\documentclass{article}
\usepackage{spconf,amsmath,graphicx,hyperref, amssymb, booktabs, enumitem}
\usepackage{xcolor}

\ninept

\title{Shortcut Flow Matching for Speech Enhancement: \\ Step-Invariant flows via single stage training}
\name{%
  \begin{tabular}{c}
    Naisong Zhou$^{\dagger \star}$\thanks{This project was completed during an internship at Logitech.},\;
    Saisamarth Rajesh Phaye$^{\star}$,\;
    Milos Cernak$^{\star}$,\;
    Tijana Stojkovi\'c$^{\star}$,\;
    Andy Pearce$^{\star}$ \\
    \textit{Andrea Cavallaro}$^{\dagger}$,\;
    \textit{Andy Harper}$^{\star}$
  \end{tabular}
}
\address{$^{\dagger}$EPFL, Lausanne, CH, $^{\star}$Logitech, Lausanne, CH}

\begin{document}
\maketitle
\begin{abstract}
Diffusion-based generative models have achieved state-of-the-art performance for perceptual quality in speech enhancement (SE). However, their iterative nature requires numerous Neural Function Evaluations (NFEs), posing a challenge for real-time applications. On the contrary, flow matching offers a more efficient alternative by learning a direct vector field, enabling high-quality synthesis in just a few steps using deterministic ordinary differential equation~(ODE) solvers. We thus introduce \textbf{S}hortcut \textbf{F}low \textbf{M}atching for \textbf{S}peech \textbf{E}nhancement (SFMSE), a novel approach that trains a single, step-invariant model. By conditioning the velocity field on the target time step during a one-stage training process, SFMSE can perform single, few, or multi-step denoising without any architectural changes or fine-tuning. Our results demonstrate that a single-step SFMSE inference achieves a real-time factor (RTF) of 0.013 on a consumer GPU while delivering perceptual quality comparable to a strong diffusion baseline requiring 60 NFEs. This work also provides an empirical analysis of the role of stochasticity in training and inference, bridging the gap between high-quality generative SE and low-latency constraints. 
\end{abstract}

\section{Introduction}
Speech enhancement (SE) must jointly optimize perceptual quality, algorithmic latency, and compute budget.  For natural conversation in interactive settings, the end-to-end delay is typically expected to remain well below a few hundred milliseconds. This places pressure on SE models to deliver high-quality denoising under tight runtime constraints.

Classical approaches such as Wiener filtering~\cite{weinerfiltering} assume relatively stationary noise, are computationally light but their performance is limited in challenging conditions. Data-driven SE has since advanced robustness by learning from large corpora. Data-driven SE methods can be grouped into \emph{predictive} and \emph{generative} models. Predictive models (e.g., Conv-tasnet~\cite{Luo2018ConvTasNet}) learn a direct mapping or masking from noisy to clean signals~\cite{wang2016masking}. Generative models (e.g., MelGAN~\cite{kumar2019melgan}) learn a conditional distribution and can generalize better when synthesis of low-level detail is required.

Diffusion models deliver strong naturalness in audio generation and have been adapted to SE in the complex STFT (short, overlapping window Fourier transform producing complex time–frequency coefficients) and waveform domains~\cite{Kong2021DiffWave,richter2023cdiffuse,Welker2022SGMSE}. 
However, the reverse process is typically computationally intensive, often requiring tens of reverse steps, each composed of 1 or 2 NFE (often when predictor-corrector is used, like in \cite{song2021scorebased}) and thus conflicting with low-latency operation. Recent work has explored acceleration, including consistency training for one/few-step mappings~\cite{Song2023Consistency}, online/buffered formulations that amortize computation over streaming windows~\cite{lay2025diffusionbufferonlinediffusionbased}, and second-stage corrections that enable few-step sampling in CRP~\cite{lay2024singlefewstepdiffusiongenerative}. %CRP particularly  the error  

Flow matching~\cite{Lipman2022FlowMatching} and rectified flow~\cite{Liu2022RectifiedFlow} learn the vector field directly,  i.e., a function $v_\theta(\mathbf{x},t,\mathbf{y})\in\mathbb{R}^d$ that maps a state and time (and conditioning) to its instantaneous velocity. The associated ordinary differential equation (ODE) evolves deterministically as $\dot{\mathbf{x}}_t = v_\theta(\mathbf{x}_t,t,\mathbf{y})$ (with an initial condition), which can be integrated with coarse-step solvers at inference. Shortcut flow matching~\cite{Frans2024Shortcut} further conditions the velocity on a target time step, enabling a single network to support single-, few-, and multi-step trajectories. Recent SE studies within conditional flow matching frameworks report competitive quality with NFEs as low as~five~\cite{lee2025flowseflowmatchingbasedspeech,wang2025flowseefficienthighqualityspeech,cross2025flowingstraighterconditionalflow}.

\textbf{Contributions.} We investigate shortcut flow matching for SE and make the following contributions:
\begin{itemize} 
  \item A shortcut-conditioned flow matching formulation for SE that exposes the target time step to the velocity field, enabling single, few, and multi-step inference with one model.
  \item An empirical study of stochasticity in training/inference and its impact on stability and quality.
  \item Real-time results: with a single NFE, our SFMSE attains quality comparable to diffusion baselines using the same backbone and representations at 60 NFEs, while achieving an RTF of 0.013 on a consumer GPU.
\end{itemize}

\section{Background}
\subsection{Task Definition}
Let $\mathbf{Y},\mathbf{S},\mathbf{N}\in\mathbb{C}^{F\times T}$ denote the complex STFTs of the noisy speech, clean speech, and noise, respectively, with an additive mixture model $\mathbf{Y = S + N}$.  $F$ denotes number of frequency bins and $T$ is that of time windows.
Flow-based (e.g., diffusion, flow matching, Schrödinger bridges) speech enhancement is posed as conditional generation, where models learn a continuous path between a simple prior distribution $p_1$ and the clean speech distribution $p_0$. 

\subsection{Diffusion and Flow Matching}
The diffusion process is described with an Stochastic Differential Equation (SDE): 
\begin{equation}
\label{eq:fwd_sde}
d\mathbf{x}_t \;=\; f(\mathbf{x}_t,t)\,dt \;+\; g(t)\,d\mathbf{w}_t,
\end{equation} 
where $f$ is the drift, $g$ the diffusion schedule, and $\mathbf{w}_t$ a standard Wiener process whose increments are Gaussian with pre-computable variance. 
Discretizing with Euler–Maruyama and a backward step $\Delta t>0$ from $t_k$ to $t_{k-1}=t_k-\Delta t$ yields
\begin{equation}
\label{eq:euler_maruyama}
\begin{aligned}
\mathbf{x}_{k-1} \;=\; \mathbf{x}_k
&\;+\; \Delta t\,\Big[f(\mathbf{x}_k,t_k) - g(t_k)^2\, s_\theta(\mathbf{x}_k,t_k,\mathbf{y})\Big] \\
&\;+\; g(t_k)\sqrt{\Delta t}\;\mathbf{z}_k,\qquad \mathbf{z}_k\sim\mathcal{N}(\mathbf{0},\mathbf{I}),
\end{aligned}
\end{equation}
showing a decomposition into a deterministic drift and a stochastic term of scale $g(t_k)\sqrt{\Delta t}$.

FM forgoes explicit score estimation and directly learns a time–dependent velocity field $v_\theta$ whose ODE transports the endpoint $p_1(\cdot\mid\mathbf{y})$ to $p_0(\cdot\mid\mathbf{y})$ \cite{Lipman2022FlowMatching,Liu2022RectifiedFlow}:
\begin{equation}
\label{eq:fm_ode}
% \begin{align}
% &\frac{d\mathbf{x}_t}{dt} \;=\; v_\theta(\mathbf{x}_t,t,\mathbf{y}),\\
\mathbf{x}_{t+d} = \mathbf{x}_d + v_{\theta}(\mathbf{x}_t,t,\mathbf{y})*d.
% \end{align}
\end{equation}
The target velocity is typically induced by interpolation; in rectified flow, a near-linear interpolation is used, which empirically improves robustness under coarse ODE stepping \cite{Liu2022RectifiedFlow}.

Comparing  Eq.~\eqref{eq:euler_maruyama} and the ODE update implied by Eq.\eqref{eq:fm_ode}, the reverse–SDE step equals the same drift term plus an additive Gaussian perturbation. Thus, at a high level the algorithmic difference is SDE (stochastic drift + noise) versus ODE (deterministic velocity). Because FM dynamics are deterministic, whether the overall generator is stochastic depends on the endpoint: with a random $p_1$ (e.g., a standard or observation–centred Gaussian) the model remains generative; with a deterministic endpoint (e.g., $\delta_{\mathbf{y}}$), it reduces to a conditional mapping. We therefore report an empirical comparison of $p_1$ choices. For diffusion (and Schrödinger–bridge formulations), observation–centred priors such as $\mathcal{N}(\mathbf{y},\sigma^2\mathbf{I})$ and their $\sigma\!\to\!0$ limit $\delta_{\mathbf{y}}$ are theoretically admissible and often behave similarly in practice.

% \subsection{Accelerated Sampling}
% % A bit introduction 
% Shortcut training \cite{Frans2024Shortcut} conditions on a step size $d$ and enforces self-consistency,
% $s_\theta(x,t,2d)\!\approx\!\tfrac12 s_\theta(x,t,d)+\tfrac12 s_\theta(x+d\,s_\theta(x,t,d),t,d)$,
% so that a single network supports many/few/one-step inference; see Sec.~\ref{sec:method} for our SE instantiation.

\section{Method}
\label{sec:method}
\subsection{Flow Matching for Speech Enhancement}
We formulate conditional speech enhancement (SE) with flow matching (FM) by learning a velocity field that transports probability mass from an endpoint prior $p_{1}(\mathbf{x}\mid\mathbf{y})$ to the clean conditional $p_{0}(\mathbf{x}\mid\mathbf{y})$ along an ODE. Let $\mathbf{y}$ be the noisy observation and $\mathbf{x}_{t}\in\mathbb{C}^{F\times K}$ the STFT-domain state at time $t\in[0,1]$. The learnable field is $v_{\theta}(\mathbf{x}_{t},t,\mathbf{y})$. This setup requires: (i) conditionality—$v_{\theta}$ depends on $\mathbf{y}$ and the current state $\mathbf{x}_{t}$; (ii) an endpoint prior $p_{1}(\cdot\mid\mathbf{y})$ that is easy to sample and compatible with enhancement; and (iii) a teacher/path design whose target velocity is stable under coarse ODE steps.

Following rectified flow matching~\cite{Liu2022RectifiedFlow}, we adopted linear interpolation for intermediate states $\mathbf{x_t}$:
\begin{equation}
  \mathbf{x_t}\;=\; (1-t)\,\mathbf{x}_{0} + t\,\mathbf{x}_{1}, 
  \qquad 
  v_{target} \;=\; \mathbf{x}_{1}-\mathbf{x}_{0}.
\end{equation}
At inference, the learned ODE is integrated backward from $p_1$ to recover $\mathbf{x}_0$ conditioned on $\mathbf{y}$, thus enhancing the signal.

Prior FlowSE variants have integrated flow matching into SE by defining conditional velocities via linear or stochastic interpolants, and typically evaluating under a fixed few-step ODE schedule. 
Our approach differs in that (i) we augment FM with step-aware shortcut conditioning (Sec.~\ref{subsec:dt-constrants}), enabling a \emph{single} model to operate at one/few/many steps without additional retraining; and (ii) we decouple stochasticity from the dynamics by placing it solely in the endpoint prior $p_1(\cdot\mid\mathbf{y})$, followed by deterministic ODE evolution. 
We further report a controlled ablation over $p_1$ choices (Sec.~\ref{sec:results}) to isolate the impact of endpoint randomness.
% sgmse-bbed-crp: mean 4.33 std 0.61 95%CI [4.29, 4.37] n=824
% sgmse-30-steps: mean 4.30 std 0.51 95%CI [4.27, 4.34] n=824
% shortcut-ncsnpp-init-as-sgmse: mean 3.93 std 0.67 95%CI [3.89, 3.98] n=824
% shortcut-ncsnpp-init-data-std: mean 3.80 std 0.76 95%CI [3.74, 3.85] n=824
% noisy_16000: mean 3.47 std 0.82 95%CI [3.42, 3.53] n=804

\subsection{Enforcing \texorpdfstring{$d t$}{dt} Constraints with Self-Consistency Losses}
\label{subsec:dt-constrants}
We follow the shortcut \cite{Frans2024Shortcut} training to enforce one-/few-/many-step capability with a single network. The network produces a step-conditioned update $f_\theta(\mathbf{x},t,d,\mathbf{y})$ that approximates the finite-difference displacement from $\mathbf{x}$ over a step of size $dt$ at time $t$. Training enforces self-consistency across step sizes by matching a single $2d$ step with two smaller sequential $dt$ steps:
\begin{equation}
\label{eq:shortcut_sc}
\begin{aligned}
f_\theta(\mathbf{x},t,2dt,\mathbf{y})
&\approx \tfrac{1}{2}\,f_\theta(\mathbf{x},t,dt,\mathbf{y}) \\
&   +      \tfrac{1}{2}\,f_\theta\!\big(\mathbf{x}+d\,f_\theta(\mathbf{x},t,dt,\mathbf{y}),\,t,\,d,\,\mathbf{y}\big).
\end{aligned}
\end{equation}
We minimize the squared deviation from~\eqref{eq:shortcut_sc} over $(\mathbf{s},\mathbf{y})$, $t$, and $d$. In the limit $d\!\to\!0$, the quantity $f_\theta(\mathbf{x},t,d,\mathbf{y})/d$ converges to a time-dependent velocity, and~\eqref{eq:shortcut_sc} reduces to a velocity-regression objective consistent with flow matching.

The sampling of $(t,dt)$ is simple and uses different examples for different targets. Following~\cite{Frans2024Shortcut}, we split each batch with ratio $ratio\_sc$: a fraction is used for self-consistency targets and the remainder for flow-matching targets. For flow-matching targets, we sample $t$ uniformly over a grid $\{k\,dt_{\min}\}$ (with $k\in\mathbb{N}^+$) and fix $dt=dt_{\min}$. For self-consistency targets, we first sample the step size from powers of two, $dt=2^{-k}$ with $k\in\mathbb{N}^{+}$, then draw admissible $(t,dt)$ pairs. Additionally, with probability $\rho\in[0,0.2]$, we map selected pairs to $(t,dt)\!\mapsto\!(0,dt)$ to emphasize the common entry state while preserving uniform coverage elsewhere. Unless noted, we use $[dt_{\min},dt_{\max}]=[1/K_{\max},\,1/2]$ and the same $(t,dt)$ distributions during training and evaluation.

Inference follows directly. For a single step ($K=1$), we take $d=1$ and compute
\begin{equation}
\mathbf{x}_0 \;=\; \mathbf{x}_1 + f_\theta(\mathbf{x}_1,\,0,\,d,\,\mathbf{y}).
\end{equation}
For $K$ steps, we apply $K$ uniform updates with $d=1/K$, updating the time argument accordingly. We report NFEs, RTF, and SE metrics for $K\in\{1,2,4,8,16\}$.

\subsection{Prior Distributions}
\label{subsec:prior-dist}

% To study the role of endpoint stochasticity, we compare three choices of $p_1(\mathbf{x}\mid\mathbf{y})$ while keeping architecture, loss, schedules, and preprocessing fixed. 

% The first is a standard Gaussian $\mathcal{N}(\mathbf{0},\mathbf{I})$, which initializes trajectories far from the observation. The second is an observation-centred Gaussian $\mathcal{N}(\mathbf{y},\sigma^2\mathbf{I})$ that starts near the mixture with controllable variance $\sigma^2$. In SGMSE \cite{Welker2022SGMSE} the prior $p1$ also follows this second way. For fair comparison, we adopt $sigma$ in prior same as in SGMSE, where the default noise scheduler function produces $\sigma(t)$ and $\sigma(0)=0.389$.  The third is the deterministic limit $\delta_{\mathbf{y}}$ obtained as $\sigma\to 0$, which removes endpoint randomness and yields a conditional mapping under the learned ODE dynamics. Despite this 3 prevailing prior, we also add an additional fourth dynamic prior where $x1\sim N(y, \alpha  \sigma(y) I) $ where the variance is based on the variance in this particular data sample and $\alpha$ is an empirical parameter 0.2. 

% In all cases, sampling proceeds by drawing $\mathbf{x}_1\sim p_1(\cdot\mid\mathbf{y})$ (or setting $\mathbf{x}_1=\mathbf{y}$ for $\delta_{\mathbf{y}}$) and applying one- or few-step updates using $f_\theta$. This ablation isolates how the prior choice influences conditioning near $t\approx 1$, one-/few-step accuracy, and runtime.

To assess the effect of endpoint stochasticity, we compare four options of $p_1(\mathbf{x}\mid\mathbf{y})$ while keeping the architecture, loss, schedules, and preprocessing fixed.

\begin{itemize}
    \item \textbf{Shortcut-G}: A standard Gaussian $\mathcal{N}(\mathbf{0},\mathbf{I})$, which initializes trajectories far from the observation.
    \item \textbf{Shortcut-S}: An observation-centered Gaussian $\mathcal{N}(\mathbf{y},\sigma^2\mathbf{I})$ with tunable variance $\sigma^2$. In SGMSE~\cite{Welker2022SGMSE}, the endpoint prior follows this form; for comparability, we use the same noise schedule, i.e., $\sigma(t)$ with $\sigma(1)=0.389$.
    \item \textbf{Shortcut-D}: A data-adaptive prior $\mathcal{N}\!\big(\mathbf{y},\,\alpha\,\sigma^2(\mathbf{y})\,\mathbf{I}\big)$, where $\widehat{\sigma}^2(\mathbf{y})$ estimates the sample-specific variance and $\alpha=0.2$ controls its magnitude.
    \item \textbf{Shortcut-F}: The deterministic limit $\delta_{\mathbf{y}}$ obtained as $\sigma\to 1$, which removes endpoint randomness and yields a purely deterministic conditional mapping under the learned ODE. This shortcut F, also ingest clean signal y as classifer free guidance \cite{ho2022classifierfreediffusionguidance}.
\end{itemize}

In all cases, we sample $\mathbf{x}_1\sim p_1$ (or set $\mathbf{x}_1=\mathbf{y}$ for $\delta_{\mathbf{y}}$) and then apply one or few-step updates using $f_\theta$ during inference. This ablation isolates how the endpoint prior influences conditioning near $t\approx 1$, one-/few-step accuracy.

% Milos
% non_causal_ncsnpp: mean 4.16 std 0.65 +- 0.04 n=824
% g-shortcut-init-gaussian_16000: mean 3.95 std 0.69 +- 0.05 n=821

\begin{table*}[t]
\centering
\caption{Performance of the proposed shortcut systems with the reference SGMSE 60 NFEs system and baseline CRP 1 NFE. All systems are trained and validated on the same VB-DMD data. The bold values show the 1 NFE statistically best metrics.}
\label{tab:metrics}
\begin{tabular}{lcccccccccc}
\toprule
 & NFE & ESTOI & SI-SDR (dB) & POLQA & OVRL-MOS & SIG-MOS & BAK-MOS & P.800 MOS & NISQA MOS\\
\midrule
\midrule
Noisy & -- & 0.79±0.11 & 8.45±4.23 &3.47±0.06 & 2.69±0.39 & 3.34±0.39 & 3.12±0.56 & 3.05±0.29  & 3.17±0.06\\
\midrule
SGMSE & 60 & 0.86±0.07 & 17.45±2.47 & 4.30±0.03 & 3.17±0.15 & 3.48±0.12 & 3.98±0.13 & 3.53±0.21  &4.63±0.02\\
\midrule
SGMSE & 1 & 0.02±0.02 & -25.32±1.28 & -- & 1.20±0.06 & 1.56±0.17 & 1.29±0.06 & 2.08±0.02 & 1.14±0.00\\
CRP & 1 & 0.84±0.09 & 18.04±3.37& \textbf{4.33±0.04} & \textbf{3.05±0.23} & \textbf{3.38±0.20} & 3.90±0.21 & \textbf{3.53±0.23}   & \textbf{4.38±0.03}\\
\midrule

Shortcut-G & 1 & 0.79±0.09 & 15.60±2.44 & 3.95±0.05 & 2.92±0.21 & 3.26±0.19 & 3.83±0.14 & 3.37±0.22 & 3.93±0.03\\
Shortcut-S & 1 & 0.83±0.08 & 16.32±2.56 &  3.93±0.04 & 3.02±0.18 & 3.37±0.15 & 3.84±0.18 & 3.45±0.21  & 3.94±0.04\\
Shortcut-D & 1 & 0.82±0.09 & 17.14±2.76  & 3.80±0.06 & 2.98±0.19 & 3.36±0.16 & 3.77±0.20 & 3.41±0.21  & 3.84±0.04\\
Shortcut-F & 1 & \textbf{0.86±0.08} & \textbf{18.39±2.92}  & 4.16±0.04 & 3.02±0.21 & 3.34±0.18 & 3.90±0.17 & 3.47±0.23 & 4.29±0.03\\

\bottomrule
\end{tabular}
\end{table*}

\begin{figure*}[t]
  \centering
  \includegraphics[width=\textwidth]{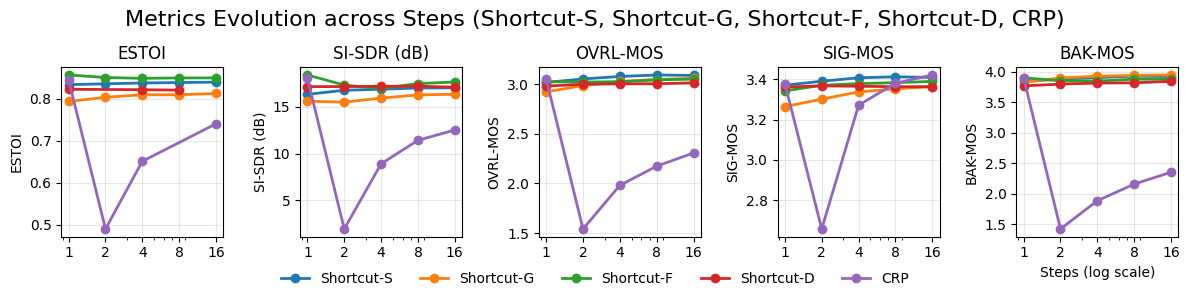}
  \caption{Metrics evolution across steps for Shortcut-S/G/F/D and CRP.}
  \label{fig:metrics-steps}
\end{figure*}

\section{Experimental setup}
\subsection{Datasets for training and evaluation}

Following prior work\cite{Welker2022SGMSE}\cite{lay2024singlefewstepdiffusiongenerative} we adopt the VoiceBank–DEMAND (VB-DMD) dataset for training and testing \cite{Valentini2017VBD}, which consists of recordings from 28 speakers and is a widely used public benchmark for single-channel speech enhancement. Noisy mixtures are created by combining clean utterances from the Voice Bank corpus with 10 DEMAND noise types at RMS SNRs of 0, 5, 10, and 15 dB. The test set uses two unseen speakers and five unseen noise types at SNRs of 2.5, 7.5, 12.5, and 17.5 dB. The durations of the training, validation, and test splits are 8 h 45 min, 37 min, and 34 min, respectively.  

\subsection{Baselines and Implementation Details}
We train shortcut flow matching with an NCSN++ v2 \cite{song2021scorebased} backbone to obtain a model size comparable to multi-step references SGMSE \cite{Welker2022SGMSE} and CRP \cite{lay2024singlefewstepdiffusiongenerative}, which we selected as our baselines. For the flow-matching component we set the minimum step size to $\mathrm{dt_{min}}=1/128$. For creating self-consistency targets, we sample with $\mathrm{rate_{sc}}=0.25$, and we weight the self-consistency term by $\lambda_{\mathrm{sc}}=0.1$. Audio preprocessing follows the original SGMSE configuration. All shortcut models are trained on VB-DMD for 100 epochs using Adam with learning rate $\eta=10^{-4}$. For longer recordings, we chunk audio into fixed-length non-overlapping chunks, enhance each chunk individually and concatenate the enhanced outputs. All evaluations are run on an NVIDIA RTX 4070Ti.

\subsection{Metrics}

We report both reference-based and reference-free metrics. 
The former compares enhanced speech against clean ground truth and is used for both in-domain and out-of-domain evaluation; the latter relies on learned predictors and does not require references. 
Among all the quality metrics, higher scores are better.

For reference-based metrics, we use POLQA \cite{beerends2013polqa} for perceptual quality (1–4.5), ESTOI \cite{taal2011estoi} for intelligibility (0–1), and SI-SDR \cite{sisdr2019} for signal fidelity (dB).

For reference-free metrics, we use DNSMOS P.835 \cite{dns2022}, reporting Speech Quality (SIG), Background Noise Quality (BAK), and Overall Quality (OVRL). We also reported NISQA MOS from \cite{Mittag_2021}. In the subjective listening test on VB-DMD, we reported the P800 MOS evaluation score\cite{recommendation1996800}.

\subsubsection{3QUEST Metrics}
In addition to the above mentioned metrics, we also include an analysis using the 3QUEST algorithm in accordance with the Microsoft Teams Certification specification for conferencing devices \cite{microsoft_teams_certification_specs}. Nineteen utterances from the IEEE sentences are replayed using a HATS system, under ACQUA control, in an ETSI compliant acoustically treated room. Following the guidelines from the Teams specification, noises commonly found in conference room situations were replaying into the room along with the speech (general meeting noise, notebook keyboard, PC keyboard, projector).  

Additionally, HVAC noise was also replayed at level of 42 dBA (normal), 49 dBA (loud), and 57 dBA (very loud) when measured at the recording device, with speech reproduced at 89 dB SPL (normal) and 79 dB SPL (quiet) at the mouth reference position, consistent with typical Teams/Zoom certification settings. Each recording was approximately 79s.

For each recording, we apply the checkpoint trained on VB-DMD for enhancement only. 

Following the constraints in Teams certification \cite{microsoft_teams_certification_specs} we report 3 metrics: 
NOBGN-SMOS which denotes the speech quality when no background noise is replayed, simulating the conversations in a well treated conference room; SMOS to denote the general speech quality in noisy conditions; and NMOS for the noise denoising quality.  The certification includes thresholds on each metric for a device to be certificated, and we include them in Table \ref{tab:certification_results} for reference.

\section{Results}

\label{sec:results}
Table \ref{tab:metrics} summarizes the performance and NFEs of the proposed SFMSE on the VB-DMD datasets. For baselines, we directly used checkpoints form SGMSE\cite{Welker2022SGMSE} and CRP was fine-tuned with $n_{rev}$=1, based on the pretrained weights in \cite{lay23_interspeech}. We report the mean and 95\% confidence interval of each measure. We can see that our shortcut models have roughly equivalent scores to SGMSE with 60 NFE. When compared with CRP, our method generally performs slightly lower, but we do not require any fine-tuning stage, and the confidence interval is smaller than CRP, indicating better robustness and generalization capability.

The 4 shortcut variants represents the 4 different priors, as introduced in Section \ref{subsec:prior-dist}.
% Among them, -G represents the Gaussian prior, -S represents the centered distribution around noisy signal with $\sigma=\sigma(1) = 0.389$ in SGMSE, -D is the data-adaptive prior $\mathcal{N}\!\big(\mathbf{y},\,\alpha\,\sigma^2(\mathbf{y})\,\mathbf{I}\big)$, -F denotes the degraded prior $\delta_{\mathbf{y}}$, which leads to a deterministic mapping from noisy speech to clean speech. 
Among the 4 variants, the centered priors (Shortcut-S / Shortcut-D) explicitly encode the belief that $\mathbf{x}_1$ should lie near the observation $\mathbf{y}$ (e.g., $\mathcal{N}(\mathbf{y}, \sigma^2 \mathbf{I})$ or $\mathcal{N}(\mathbf{y}, \alpha\,\sigma^2(\mathbf{y})\,\mathbf{I})$), which shortens the transport path and reduces target variance; under the extremely low-step regime (NFE=1), this lower-variance target is easier to learn and thus more stable than a pure Gaussian prior. In contrast, the uncentered Gaussian prior initiates far from $\mathbf{y}$, injecting task-irrelevant randomness and increasing bias/variance for a single-step predictor, hence its weaker performance. 

Interestingly, the deterministic endpoint prior (Shortcut-F, $\delta_{\mathbf{y}}$) collapses endpoint stochasticity and produces the sharpest one-step estimates in-distribution (best means across several metrics). As this is trained and tested on the same datase, we explore with the unseen 3QUEST dataset whether by excluding uncertainty modeling it Shortcut -F would generalize worse under distribution shift.

Furthermore, we have shown in Figure\ref{fig:metrics-steps} the scores evolving with NFE $\in \{1,2,4,8,16\}$ for the proposed model, with CRP model tuned on $N_{rev}$ mentioned above. We observe that, in the single-step case, shortcut variants have similar performance with CRP, and when the number of steps used increases, shortcut can still have the same level of performance without extra fine-tuning. With just a single-stage joint training, the model achieves comparable enhancement quality to upper bound in \cite{Welker2022SGMSE} with fewer steps at different scale.

\begin{table}
\small
\caption{Certification Results (Step=1) for the 3QUEST test}
\label{tab:certification_results}
\begin{tabular}{lcccc}
\toprule
 & NoBGN-SMOS & SMOS & NMOS \\
\midrule
% g-vanilla-fm-ncsnpp & 4.06 & 3.97 & 3.76 \\
% @naisong you can add a citation here
Teams Threshold \cite{microsoft_teams_certification_specs}  & 4.0 & 3.50 & 2.90 \\
Shortcut-F & 4.16 & 4.09 & 3.69\\
Shortcut-S & 4.16 & 4.03 & 3.78\\ 
Shortcut-D & 4.05 & 3.87 & 3.82\\
Shortcut-G & 3.85 & 3.71 & 3.35\\
\bottomrule
\end{tabular}
\end{table}

\begin{figure}[t]
  \centering
  \includegraphics[width=0.9\linewidth]{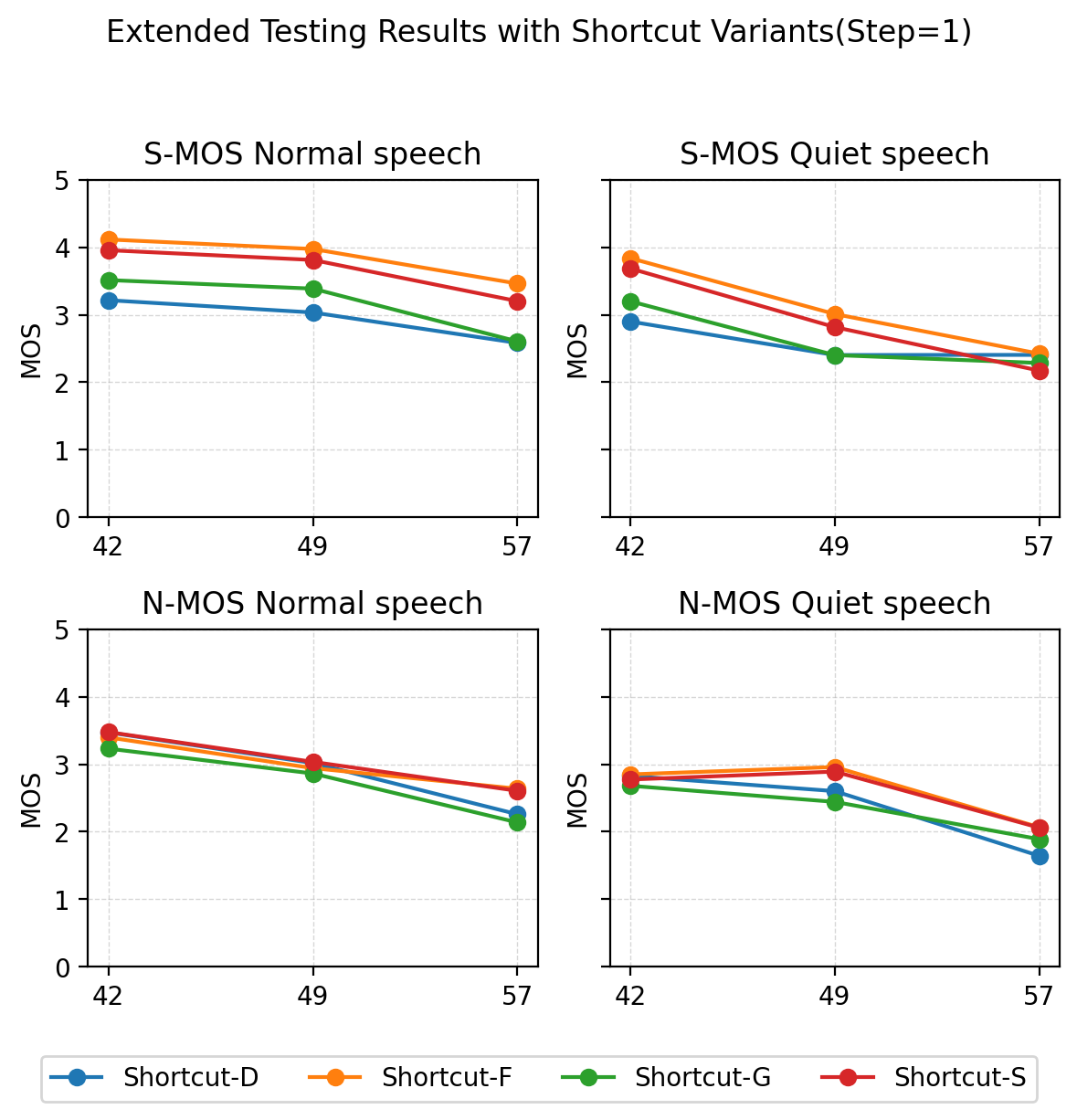}
  \caption{Extended testing results by variant for step=1. 
  Each subplot shows MOS under different SNR levels (42, 49, 57 dBA) 
  for Normal and Quiet speech, across both S-MOS and N-MOS conditions.}
  \label{fig:extended-results}
\end{figure}

Table~\ref{tab:certification_results} and Figure~\ref{fig:extended-results} present the 3QUEST results on the additional dataset. As shown in Table~\ref{tab:certification_results}, all shortcut variants—except for Shortcut-G—meet the Teams threshold for speech and noise cancellation quality under noisy or good conference conditions. Shortcut-G falls slightly below for NoBGN-SMOS, suggesting that it may be over-suppressing parts of the audio. Figure~\ref{fig:extended-results} further indicates that the shortcut models remain competitive under normal and loud HVAC noise conditions, especially when with normal speech. At very high noise levels (57 dBA), however, performance degrades. This is partly due to the recording pipeline: an automatic gain control (AGC) stage, which cannot be disabled during evaluation, tends to misclassify quiet speech segments as noise and attenuate them. Consequently, the input waveforms at 57dBA are already incomplete before enhancement.

\section{Conclusion}
We integrated Shortcut Flow Matching into speech enhancement. Our SFMSE model (trained and evaluated on VB-DEMAND) with NFE=1 achieves enhancement quality comparable to SGMSE at NFE=60. Relative to other one-step or few-step methods, SFMSE eliminates two-stage fine-tuning and achieves step-invariant performance through a single joint-training stage, reducing computational cost and design complexity. By comparing alternative priors, we analyzed the role of endpoint stochasticity in flow matching and its effect on enhancement quality, and found that a deterministic prior does not degrade generalization on the unseen 3QUEST dataset. On a consumer-grade GPU, the system operates with an RTF of 0.013, satisfying real-time requirements. 

However, SFMSE 1-step quality does not surpass a 1-step fine-tuned CRP model. Future work targets improving overall quality with richer conditioning (e.g., noise/context embeddings) and more precise objectives (e.g., multi-resolution STFT/perceptual losses), as well as streamable deployment and evaluation under causality.

The authors would like to thank Yuliia Shandra for their valuable feedback and insights, and the anonymous listeners for their assistance.

\vfill \pagebreak

\bibliographystyle{IEEEbib}
\bibliography{refs}

\begin{thebibliography}{10}

\bibitem{weinerfiltering}
W.A. Gardner,
\newblock ``{{Cyclic Wiener filtering}}: theory and method,''
\newblock {\em IEEE Transactions on Communications}, vol. 41, no. 1, pp. 151--163, 1993.

\bibitem{Luo2018ConvTasNet}
Yi~Luo and Nima Mesgarani,
\newblock ``{Conv-TasNet: Surpassing Ideal Time-Frequency Magnitude Masking for Speech Separation},''
\newblock {\em arXiv preprint arXiv:1809.07454}, 2018.

\bibitem{wang2016masking}
Donald~S. Williamson, Yuxuan Wang, and DeLiang Wang,
\newblock ``{Complex Ratio Masking for Monaural Speech Separation},''
\newblock {\em IEEE/ACM Transactions on Audio, Speech, and Language Processing}, vol. 24, no. 3, pp. 483--492, 2016.

\bibitem{kumar2019melgan}
Kundan Kumar et~al.,
\newblock ``{MelGAN: Generative adversarial networks for conditional waveform synthesis},''
\newblock in {\em {NeurIPS}}, 2019.

\bibitem{Kong2021DiffWave}
Zhifeng Kong, Wei Ping, Jiaji Huang, Kexin Zhao, and Bryan Catanzaro,
\newblock ``{DiffWave: A Versatile Diffusion Model for Audio Synthesis},''
\newblock in {\em {ICLR}}, 2021.

\bibitem{richter2023cdiffuse}
Julius Richter et~al.,
\newblock ``{CDiffuSE: Conditional Diffusion for Speech Enhancement},''
\newblock in {\em {Interspeech}}, 2023.

\bibitem{Welker2022SGMSE}
Simon Welker, Julius Richter, and Timo Gerkmann,
\newblock ``{Speech Enhancement with Score-Based Generative Models in the Complex STFT Domain},''
\newblock in {\em {Interspeech}}, 2022.

\bibitem{song2021scorebased}
Yang Song, Jascha Sohl-Dickstein, Diederik~P Kingma, Abhishek Kumar, Stefano Ermon, and Ben Poole,
\newblock ``{Score-Based Generative Modeling through Stochastic Differential Equations},''
\newblock in {\em {International Conference on Learning Representations}}, 2021.

\bibitem{Song2023Consistency}
Yang Song, Prafulla Dhariwal, Mark Chen, and Ilya Sutskever,
\newblock ``{Consistency Models},''
\newblock {\em arXiv preprint arXiv:2303.01469}, 2023.

\bibitem{lay2025diffusionbufferonlinediffusionbased}
Bunlong Lay, Rostislav Makarov, and Timo Gerkmann,
\newblock ``{Diffusion Buffer: Online Diffusion-based Speech Enhancement with Sub-Second Latency},'' 2025.

\bibitem{lay2024singlefewstepdiffusiongenerative}
Bunlong Lay, Jean-Marie Lemercier, Julius Richter, and Timo Gerkmann,
\newblock ``{Single and Few-step Diffusion for Generative Speech Enhancement},'' 2024.

\bibitem{Lipman2022FlowMatching}
Yaron Lipman, Ricky T.~Q. Chen, Heli Ben-Hamu, Maximilian Nickel, and Matt Le,
\newblock ``{Flow Matching for Generative Modeling},''
\newblock {\em arXiv preprint arXiv:2210.02747}, 2022.

\bibitem{Liu2022RectifiedFlow}
Xingchao Liu, Chengyue Gong, and Qiang Liu,
\newblock ``{Flow Straight and Fast: Learning to Generate and Transfer Data with Rectified Flow},''
\newblock {\em arXiv preprint arXiv:2209.03003}, 2022.

\bibitem{Frans2024Shortcut}
Kevin Frans, Danijar Hafner, Sergey Levine, and Pieter Abbeel,
\newblock ``{One Step Diffusion via Shortcut Models},''
\newblock {\em arXiv preprint arXiv:2410.12557}, 2024.

\bibitem{lee2025flowseflowmatchingbasedspeech}
Seonggyu Lee, Sein Cheong, Sangwook Han, and Jong~Won Shin,
\newblock ``{FlowSE: Flow Matching-based Speech Enhancement},'' 2025.

\bibitem{wang2025flowseefficienthighqualityspeech}
Ziqian Wang, Zikai Liu, Xinfa Zhu, Yike Zhu, Mingshuai Liu, Jun Chen, Longshuai Xiao, Chao Weng, and Lei Xie,
\newblock ``{FlowSE: Efficient and High-Quality Speech Enhancement via Flow Matching},'' 2025.

\bibitem{cross2025flowingstraighterconditionalflow}
Mattias Cross and Anton Ragni,
\newblock ``{Flowing Straighter with Conditional Flow Matching for Accurate Speech Enhancement},'' 2025.

\bibitem{ho2022classifierfreediffusionguidance}
Jonathan Ho and Tim Salimans,
\newblock ``Classifier-free diffusion guidance,'' 2022.

\bibitem{Valentini2017VBD}
Cassia Valentini-Botinhao, Xin Wang, Shinji Takaki, and Junichi Yamagishi,
\newblock ``{Noisy speech database for training speech enhancement algorithms and TTS models (VoiceBank-DEMAND)},'' 2017,
\newblock Clean/noisy parallel data; 28spk/56spk splits.

\bibitem{beerends2013polqa}
John Beerends et~al.,
\newblock ``{Perceptual Objective Listening Quality Assessment (POLQA)},''
\newblock {\em Journal of the Audio Engineering Society}, 2013.

\bibitem{taal2011estoi}
Cees Taal, Richard Hendriks, et~al.,
\newblock ``{An evaluation of objective measures for intelligibility prediction of time–frequency weighted noisy speech},''
\newblock in {\em {Interspeech}}, 2011.

\bibitem{sisdr2019}
Jonathan~Le Roux, Scott Wisdom, Hakan Erdogan, and John~R. Hershey,
\newblock ``{SDR – Half-baked or Well Done?},''
\newblock in {\em {ICASSP 2019 - 2019 IEEE International Conference on Acoustics, Speech and Signal Processing (ICASSP)}}, 2019, pp. 626--630.

\bibitem{dns2022}
Chandan K~A Reddy, Vishak Gopal, and Ross Cutler,
\newblock ``{Dnsmos P.835: A Non-Intrusive Perceptual Objective Speech Quality Metric to Evaluate Noise Suppressors},''
\newblock in {\em {ICASSP 2022 - 2022 IEEE International Conference on Acoustics, Speech and Signal Processing (ICASSP)}}, 2022, pp. 886--890.

\bibitem{Mittag_2021}
Gabriel Mittag, Babak Naderi, Assmaa Chehadi, and Sebastian Möller,
\newblock ``{NISQA: A Deep CNN-Self-Attention Model for Multidimensional Speech Quality Prediction with Crowdsourced Datasets},''
\newblock in {\em {Interspeech 2021}}. Aug. 2021, ISCA.

\bibitem{recommendation1996800}
ITUTP Recommendation,
\newblock ``{P800, Methods for subjective determination of transmission quality, ITU-T},'' 1996.

\bibitem{microsoft_teams_certification_specs}
Microsoft,
\newblock ``{Microsoft Teams V5 Certifications(Audio)},'' \url{https://learn.microsoft.com/en-us/microsoftteams/devices/certification-specifications}, 2025,
\newblock Accessed: 2025-09-18.

\bibitem{lay23_interspeech}
Bunlong Lay, Simon Welker, Julius Richter, and Timo Gerkmann,
\newblock ``{Reducing the Prior Mismatch of Stochastic Differential Equations for Diffusion-based Speech Enhancement},''
\newblock in {\em {Interspeech 2023}}, 2023, pp. 3809--3813.

\end{thebibliography}

\end{document}